\documentstyle[12pt]{article}
\newcommand{\beq}{\begin{equation}}
\newcommand{\eeq}{\end{equation}}
\newcommand{\beqn}{\begin{eqnarray}}
\newcommand{\eeqn}{\end{eqnarray}}

\newcommand{\stackm}{\stackrel{\scriptstyle <}{{ }_{\sim}}} 
\begin{document}
\thispagestyle{empty}

\def\prepnum{UAB-FT-423}
\def\data{August, 1997}
\def\hepnum{hep-ph/9708494}

\begin{flushright}
{\parbox{3.5cm}{
\prepnum

\data

\hepnum
}}
\end{flushright}

\vspace{3cm}
\begin{center}
\begin{large}
\begin{bf}
LOOKING FOR QUANTUM SUSY SIGNATURES IN TOP QUARK DECAYS
AT HADRON COLLIDERS
\\
\end{bf}
\end{large}
\vspace{1cm}
Joan SOL\`A\footnote {Invited talk presented at the XVI International
Workshop on Weak Interaction and Neutrinos\, (WIN 97), Capri, Italy, 
22-28 June, 1997. To appear in the Proceedings.}\\

\vspace{0.25cm} 
Grup de F\'{\i}sica Te\`orica\\ 
and\\ 
Institut de F\'\i sica d'Altes Energies\\ 
\vspace{0.25cm} 
Universitat Aut\`onoma de Barcelona\\
08193 Bellaterra (Barcelona), Catalonia, Spain\\
\end{center}
\vspace{0.3cm}
\hyphenation{super-symme-tric sig-ni-fi-cant-ly ge-ne-ral}
\hyphenation{com-pe-ti-ti-ve}
\hyphenation{mo-dels}
\begin{center}
{\bf ABSTRACT}
\end{center}
\begin{quotation}
\noindent
\hyphenation{ob-ser-va-bles}
\noindent
We discuss the supersymmetric quantum effects on  
top quark decays within the MSSM. 
It turns out that $t\rightarrow H^+\,b$ is the most promising
candidate for carrying large  quantum SUSY signatures. 
As a result, the recent $(\tan\beta, M_{H^{\pm}})$ exclusion plots 
presented by the CDF Collaboration should be thoroughly revised in the
light of the MSSM.
\end{quotation}
 
\baselineskip=6.5mm  
 
\newpage

In this talk I propose to dwell on the supersymmetric phenomenology of
top quark decays with an eye on future machine developments such as the
upgrade of the Tevatron and the advent of the LHC.
In the absence of direct sparticle production, one naturally looks
for ``quantum signatures'' of the new physics by means of the 
indirect method of high precision measurements. 
The Minimal Supersymmetric extension of the Standard Model (MSSM) remains 
immaculately consistent with all known high precision experiments
at a level comparable to the SM\,\cite{WdeBoer}. This fact alone,
if we bare in mind the vast amount of high precision data available 
both from low-energy and high-energy physics, 
should justify all efforts to search for SUSY in present day 
particle accelerators. In this respect we wish to stress here the
possibility of seeing large virtual effects of SUSY through the interplay
between top quark and Higgs boson dynamics at hadron colliders. The typical
size of the effects that we are referring to is in general much larger
than the tiny few per cent level corrections predicted in all canonical 
gauge boson observables at LEP.

To start with, we recall that the supersymmetric
strong (SUSY-QCD) and the supersymmetric electroweak (SUSY-EW) 
corrections to the standard top quark decay,
$t\rightarrow W^+\,b$,
are well understood\,\cite{GJSH}, but unfortunately they are not 
too large --  as typically expected of gauge boson interactions. 
In the on-shell $G_F$-scheme, which is characterized by the set of inputs
$(G_F, M_W,M_Z,m_f,M_{SUSY},...)$,
they are negative and of the order of a few per cent (except in some
unlikely cases\,\cite{GJSH}). Therefore, they
approximately cancel out
against the positive SM contributions of the same order of magnitude 
and leaving the ordinary QCD effects  
($\simeq -10\%$) as the net MSSM corrections. Hence no significant
imprint of underlying SUSY dynamics is left on
$\Gamma(t\rightarrow W^+\,b)$, 
and we are led to examine other top quark decays.

Among the relevant MSSM top quark decays carrying an interesting
SUSY signature, the following two-body modes stand out:
\beqn
{\rm i}) &&t\rightarrow \tilde{t}_i\,\chi^0_{\alpha},\nonumber\\
{\rm ii}) &&t\rightarrow \tilde{b}_i\,\chi^+_{\alpha},\nonumber\\
{\rm iii}) &&t\rightarrow \tilde{t}_i\,\tilde{g},\nonumber\\ 
{\rm iv}) &&t\rightarrow H^+\,b\,.
\label{eq:decays}
\eeqn
Therein, $\tilde{t}_{i}$, $\tilde{b}_{i}$, $\chi^+_i$,
$\chi^0_{\alpha}$, $\tilde{g}$ ($i=1,2;\,\alpha=1,2,...,4$)
denote stop, sbottom, chargino, neutralino and gluino sparticles,
respectively. 
(Also quite a few three-body decays are possible and have been 
studied\,\cite{Guasch2}.)

While the first three decays in (\ref{eq:decays}) already
carry a direct SUSY signature, the
third one is meant to involve the charged Higgs boson of the MSSM
and it could bring along a significant quantum SUSY signature.  
In general the direct SUSY decays i)-iii) may also require
a higher order treatment, the reason being that some of
the final state signatures, after the sparticles have decayed into 
conventional particles and the LSP (typically the lightest neutralino
$\chi_1^0$), they may well mimic the standard top
quark decay.
For example, decay i) may lead to a signature similar to
the standard top quark decay into the final states 
$b\,l^+\,\nu$ or $b+2\,{\rm jets}$. For, the stop could
decay into $\chi^+_i\,b$, and subsequently yield the chain
$\chi^+_i\rightarrow\chi^0_1\,W^*
\rightarrow\chi^0_1\,l^+\,\nu$ or $\chi^0_1+2\,{\rm jets}$. 

Therefore, a detailed treatment of these direct SUSY modes
is in principle desirable to help disentangling the nature of the 
complicated final configurations and to enable a reliable determination
of the top quark cross-section within the MSSM.
Barring a light gluino window, which is nowadays harder and harder
to maintain, current limits on squark and gluino masses already rule out
decay iii) and most likely also decay ii). Moreover,
the typical size of the corrections to
the process i) is not too significant  (at the ten
per cent level at most\,\cite{Djouadi}). While this would amply suffice in
a high precision machine such as LEP, however for measurements to be 
performed in a hadron environment it is probably not enough to be detected.
In contrast, decay iv) may receive spectacularly large SUSY quantum 
corrections, namely of the order of
$50\%$, which certainly could not be missed -- if SUSY is there at all. 
For this reason, we are going to focus on that decay. To be sure,
$t\rightarrow H^+\,b$ has been object of many 
studies in the past\footnote{See \cite{CGGJS} and references therein.},
mainly within the context of general two-Higgs-doublet 
models ($2HDM$), and 
it is being thoroughly scrutinized in recent analyses at the 
Tevatron\,\cite{CDF}. Notwithstanding, no systematic treatment of the
MSSM quantum effects existed in the literature until very
recently\,\cite{CGGJS,Guasch4}.

The basic free parameters of our
analysis concerning the electroweak sector are contained in the
stop and sbottom mass matrices ($q=t,b$):
\begin{equation}
{\cal M}_{\tilde{q}}^2 =\left(\begin{array}{cc}
 {\cal M}_{11}^2 & {\cal M}_{12}^2 
\\ {\cal M}_{12}^2 &{\cal M}_{22}^2 \,.
\end{array} \right)\,,
\label{eq:stopmatrix}
\end{equation}
with 
\begin{eqnarray}
{\cal M}_{11}^2 &=&M_{\tilde{q}_L}^2+m_q^2\nonumber\\
&+&\cos{2\beta}(T^3_q-Q_q\,\sin^2\theta_W)\,M_Z^2\,,\nonumber\\
{\cal M}_{22}^2 &=&M_{\tilde{q}_R}^2+m_q^2\nonumber\\
&+&Q_q\,\cos{2\beta}\,\sin^2\theta_W\,M_Z^2\,,\nonumber\\
{\cal M}_{12}^2 &=&m_q\, M_{LR}^q\,,\nonumber\\
M_{LR}^{\{t,b\}}&=&A_{\{t,b\}}-\mu\{\cot\beta,\tan\beta\}\,. \ \ \ \
\end{eqnarray}
We denote by $m_{\tilde{t}_1}$
and $m_{\tilde{b}_1}$ the lightest stop and sbottom
masses. 

Crucial in the treatment of the electroweak SUSY effects is the
definition of $\tan\beta$ beyond the tree-level.
Following Ref.\cite{CGGJS} we define it by
means of the $\tau$-lepton decay of $H^\pm$:
\beq
\Gamma(H^{+}\rightarrow\tau^{+}\nu_{\tau})=
{\alpha m_{\tau}^2\,M_H\over 8 M_W^2 s_W^2}\,\tan^2\beta\,. 
\label{eq:tbetainput}
\eeq
This definition generates a counterterm
\beqn
{\delta\tan\beta\over \tan\beta}
&=&\frac{1}{2}\left(
\frac{\delta M_W^2}{M_W^2}-\frac{\delta g^2}{g^2}\right)
-\frac{1}{2}\delta Z_H\nonumber\\
&+&\cot\beta\, \delta Z_{HW}+ 
\Delta_{\tau}\,.
\label{eq:deltabeta}
\eeqn    
Notice that $\Delta_{\tau}$ above 
stands for the complete set of
MSSM one-loop effects on the $\tau$-lepton decay of $H^\pm$;
$\delta Z_{H}$ and $\delta Z_{HW}$ stand
respectively for the charged Higgs and mixed
$H-W$ wave-function renormalization factors; and the remaining 
counterterms $\delta g^2$ and $\delta M_W$ are the standard 
ones in the on-shell scheme\,\cite{BSH}.

The results are conveniently cast in terms of the relative 
correction with respect to the corresponding tree-level width, $\Gamma_0$:
\beq
\delta_{MSSM}={\Gamma_{MSSM} (t\rightarrow H^+\,b)
-\Gamma_0(t\rightarrow H^+\,b)\over \Gamma_0(t\rightarrow H^+\,b)}\,.
\label{eq:delta}
\eeq
We will present the numerical results for this quantity in the 
on-shell $\alpha$-scheme: 
\beq
(\alpha, M_W,M_Z,m_f,M_{SUSY},...)\,.
\eeq
The corresponding
results in the $G_F$-scheme are just
$\delta_{MSSM}-(\Delta r)_{MSSM}$\,\cite{GJSH}.
As it turns out that $\delta_{MSSM}>>(\Delta r)_{MSSM}$\,\cite{Garcia},  
the difference between the two schemes is not material 
in this case, i.e. the bulk of the
effect is already contained in the $\alpha$-parametrization.

A fundamental parameter to be numerically tested is $\tan\beta$. 
It is involved in the basic interaction Lagrangian for our decay:
\beq
{\cal L}_{Htb}={g\over\sqrt{2}M_W}\,H^+\,\bar{t}\,
[m_t\cot\beta\,P_L + m_b\tan\beta\,P_R]\,b+{\rm h.c.}\,,
\label{eq:LtbH}
\eeq 
where $P_{L,R}=1/2(1\mp\gamma_5)$ are the chiral projector operators.
Furthermore, $\tan\beta$ in
supersymmetric theories, like the MSSM, enters the
top and bottom quark Yukawa couplings
of the superpotential
through $1/\sin\beta$ and $1/\cos\beta$, respectively:
\beq
h_t={g\,m_t\over \sqrt{2}\,M_W\,\sin{\beta}}\,,\
h_b={g\,m_b\over \sqrt{2}\,M_W\,\cos{\beta}}\,.
\eeq
Notice that the bottom-quark Yukawa coupling may counterbalance the
smallness of the bottom mass at the expense of a large value of
$\tan\beta$. 
For a typical choice of parameters,
in Fig.1a we plot the various contributions to
(\ref{eq:delta}) from SUSY-QCD, SUSY-EW and the MSSM Higgs sector.
We also show the standard QCD correction.
The full MSSM correction is defined to be the sum of all these
individual contributions. In Fig.1b we display the evolution of
the different corrections with $m_{\tilde{b}_1}$; this is a
critical parameter governing the size of the leading (SUSY-QCD)
corrections. Indeed, the decoupling with the gluino mass is
much slower\,\cite{CGGJS}.
Still, even when $m_{\tilde{b}_1}$ is very large,
there remains an undamped SUSY-EW component (essentially controlled by
$m_{\tilde{t}_1}$) which can be sizeable enough for stop masses
in the few hundred $GeV$.
The corrections also increase with $A_t$ and $|\mu|$, and change
sign with $\mu$.  Of course, $\delta_{MSSM}\rightarrow 0$
when all sparticle masses increase simultaneously, for the
MSSM naturally decouples in the
limit $M_{SUSY}\rightarrow\infty$.

The definition (\ref{eq:tbetainput}) 
of $\tan\beta$ allows to renormalize the $H^{\pm}\,t\,b$-vertex 
in perhaps the most convenient way to deal with our main decay iv). 
Indeed, from the practical point of view, we should
recall the excellent methods for $\tau$-identification 
developed by the Tevatron
collaborations and recently used by CDF to study the 
existence region of the decay iv) in
the $(\tan\beta,M_H)$-plane\,\cite{CDF}.
However, we wish to show that this analysis may undergo dramatic 
changes when we incorporate the MSSM
quantum effects\,\cite{Guasch4}. Although CDF utilizes
inclusive $\tau$-lepton tagging, for our purposes it will
suffice to focus on the exclusive final state
$(l,\tau)$, with $l$ a light
lepton, as a means for detecting 
an excess of $\tau$-events\,\cite{DPRoy}.
To be precise, we are interested in the
$t\,\bar{t}$ cross-section leading to the decay sequences  
$t\,\bar{t}\rightarrow H^+\,b,W^-\,\bar{b}$ and 
$H^+\rightarrow \tau^+\,\nu_{\tau}$, $W^-\rightarrow l\,\bar{\nu}_l$, 
and {\it vice versa}. From the non-observation of these events,
in Figs.1c and 1d we derive the ($95\%$ C.L.) excluded regions for
$\mu<0$ and $\mu>0$, respectively. (In the latter case we choose a
heavier SUSY spectrum in order that (\ref{eq:delta}) remains
perturbative.)
Shown are the tree-level, standard
QCD-corrected
and fully MSSM-corrected results.
From inspection of these figures
it can hardly be overemphasized that the MSSM quantum effects
can be dramatic. In particular, while for $\mu<0$ the MSSM-corrected
curve is significantly more restrictive than the QCD-corrected one, 
for $\mu>0$ the bound essentially disappears from the perturbative
region ($\tan\beta\stackm 60$). 
The lesson to be learnt should be highly instructive: 
In contrast to the tiny corrections to gauge boson observables, the MSSM
quantum effects on top-Higgs boson physics can be rather large and
should not be ignored in future searches at the Tevatron and 
at the LHC.

{\bf Acknowledgements}:

\noindent
The author is thankful to the organizers of WIN 97 for their kind
invitation, and to 
J.A. Coarasa, D. Garcia, J. Guasch and R.A. Jim\'enez for helping
to prepare this contribution.  This work
has been partially supported by CICYT under project No. AEN93-0474.



\vspace{0.5cm}
\begin{center}
\begin{Large}
{\bf Figure Captions}
\end{Large}
\end{center}
\begin{itemize}
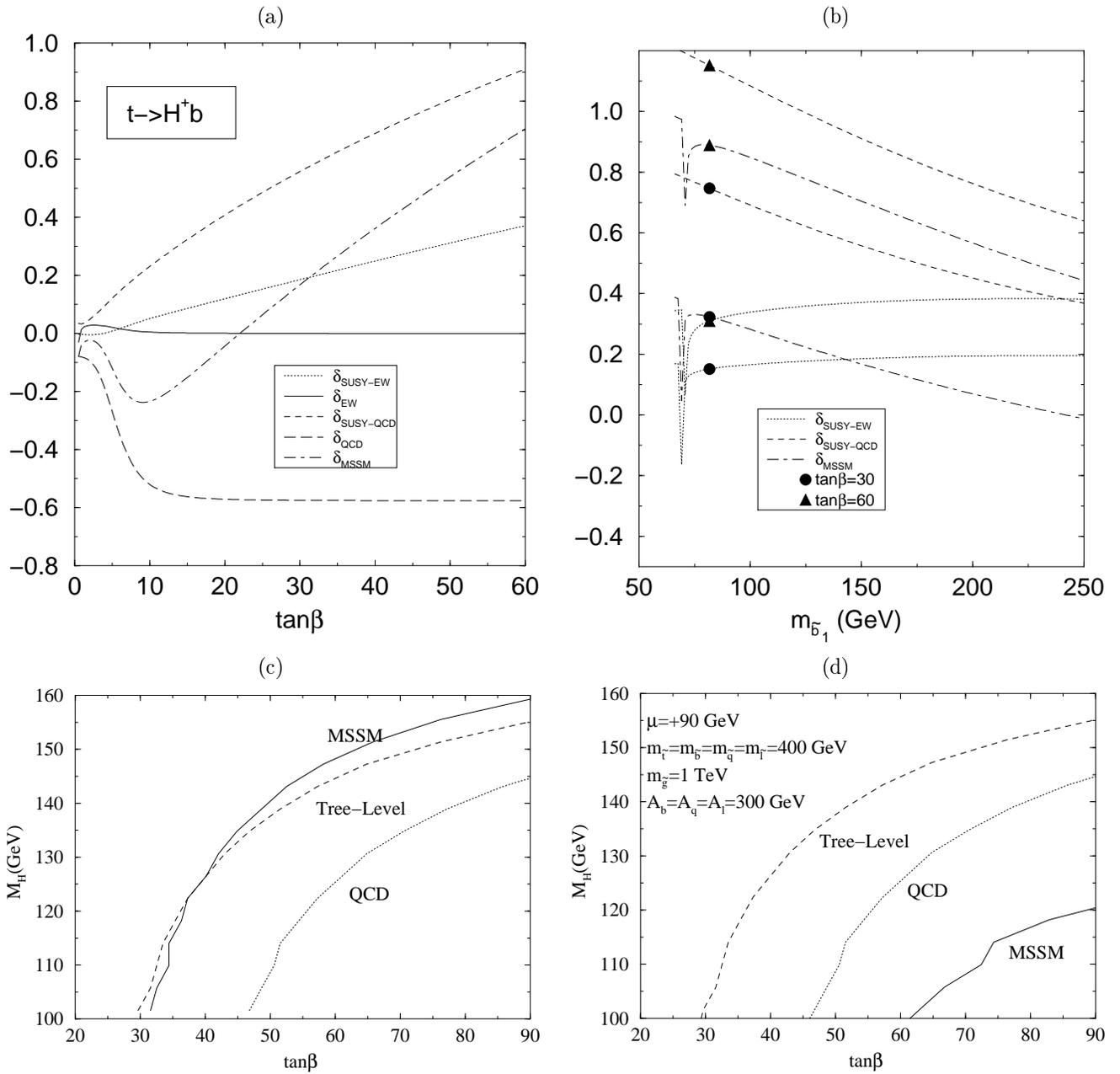
\item{\bf Fig.1}  {\bf (a)} The various individual and total
MSSM correction eq.(\ref{eq:delta})
as a function of $\tan\beta$ and given values of the
other parameters: $M_{H^{\pm}}=120\,GeV$, $\mu=-150\,GeV$, 
$m_{\tilde{g}}=300\,GeV$,
$m_{\tilde{b}_1}=150\,GeV$, $m_{\tilde{t}_1}=100\,GeV$,
$A_t=A_b=300\,GeV$;
{\bf (b)} As in (a), but as a function of $m_{\tilde{b}_1}$ and two fixed
$\tan\beta$ values;
{\bf (c)} The $95\%$ C.L. exclusion plot in the $(\tan\beta, M_{H^\pm})$-plane
for $\mu=-90\,GeV$ and remaining parameters similar to (a).
Shown are the tree-level (dashed), QCD-corrected
(dotted) and fully 
MSSM-corrected (continuous) contour lines.
The excluded region in each case is the one lying below the curve;
{\bf (d)} As in (c), but for a $\mu>0$ scenario characterized by a 
heavier SUSY spectrum.
\end{itemize}

\end{document}